\documentclass[twocolumn,showpacs,prl,superscriptaddress,floatfix]{revtex4}

\usepackage{color}
\usepackage{tabularx}
\usepackage{epsfig}
\usepackage{amsmath}
\usepackage{amssymb}
\usepackage{graphicx}
\usepackage{wasysym}
\usepackage{dcolumn}
\usepackage{bm}
\usepackage{subfigure}
\usepackage{psfrag}
\usepackage{subfigure}
\usepackage{times}

\begin{document}

\title{Efficient Cluster Algorithm for Spin Glasses in Any Space Dimension}

\author{Zheng Zhu}
\author{Andrew J.~Ochoa}
\affiliation{Department of Physics and Astronomy, Texas A\&M University,
College Station, Texas 77843-4242, USA}

\author{Helmut G.~Katzgraber}
\affiliation{Department of Physics and Astronomy, Texas A\&M University,
College Station, Texas 77843-4242, USA}
\affiliation{Materials Science and Engineering Program, Texas A\&M
University, College Station, Texas 77843, USA}
\affiliation{Santa Fe Institute, 1399 Hyde Park Road, Santa Fe, New
Mexico 87501 USA}

\date{\today}

\begin{abstract}

Spin systems with frustration and disorder are notoriously difficult to
study both analytically and numerically. While the simulation of
ferromagnetic statistical mechanical models benefits greatly from
cluster algorithms, these accelerated dynamics methods remain
elusive for generic spin-glass-like systems. Here we present a cluster
algorithm for Ising spin glasses that works in any space dimension and
speeds up thermalization by at least one order of magnitude at temperatures
where thermalization is typically difficult. Our isoenergetic cluster
moves are based on the Houdayer cluster algorithm for two-dimensional
spin glasses and lead to a speedup over conventional state-of-the-art
methods that increases with the system size. We illustrate the benefits
of the isoenergetic cluster moves in two and three space dimensions, as
well as the nonplanar chimera topology found in the D-Wave Inc.~quantum
annealing machine.

\end{abstract}


\maketitle

A plethora of problems across disciplines map onto spin-glass-like
Hamiltonians \cite{stein:13}. Despite decades of intense analytical and
numerical scrutiny, a deep understanding of these paradigmatic models of
disordered systems remains elusive.  Given the inherent difficulties of
studying these Hamiltonians analytically beyond mean-field theory as
well as the continuous increase of computer power, progress in this field
has benefited noticeably from numerical studies. The development of
efficient Monte Carlo methods such as parallel tempering
\cite{hukushima:96} and population annealing \cite{machta:10} has helped
in understanding these systems at a much deeper level; however, most
numerical studies are still plagued by corrections to finite-size
scaling due to the small system sizes currently available
\cite{hohenberg:77}.

In contrast, simulations of spin Hamiltonians without disorder and
frustration are comparably simple: Ferromagnetic systems have greatly
benefited from the development of cluster algorithms
\cite{swendsen:87,wolff:89} that help in overcoming critical slowing
down close to phase transitions.  Therefore, the holy grail of
spin-glass simulations is to introduce accelerated cluster dynamics that
improve upon the benefits of efficient simulation methods such as
population annealing or parallel tempering Monte Carlo.  In 2001
Houdayer introduced a seminal rejection-free cluster algorithm tailored
to work for two-dimensional Ising spin glasses \cite{houdayer:01}.  The
method updates large patches of spins at once, therefore effectively
randomizing the configurations and efficiently overcoming large barriers
in the free-energy landscape. Furthermore, the energy of the system
remains unchanged when performing a cluster move. This means that the
numerical overhead is very small because the rejection rate is zero and
there is no need to, for example, compute any random numbers for a
cluster update.  The use of these cluster moves made it possible to
obtain a speedup of several orders of magnitude in two-dimensional
systems, therefore allowing us to simulate considerably larger system
sizes.

While cluster algorithms such as the Swendsen-Wang and Wolff ones
\cite{swendsen:87,wolff:89} work well for ferromagnetic systems in any
space dimension because the clusters reflect the spin correlations in
the system, this is not the case for algorithms that build clusters like
the Houdayer cluster algorithm. In this case, the clusters do not
reflect overlap correlations \cite{chayes:98,machta:07} and cluster
updates only have an accelerating effect on the dynamics if the clusters
do not span the entire system or if the comprise single spins.  This is
the case either when temperatures are close to zero (small clusters) or
when the underlying geometry of the problem has a percolation threshold
below 50\%---as is the case in three space dimensions. Updating such a
system-spanning cluster amounts to swapping out both replicas, thereby
not randomizing the configurations.  This means that while the method
works in principle, it does not really provide any simulational benefit.
As such, Houdayer cluster moves work, in principle, only for models
where the percolation threshold is above 50\%, as is the case in
two-dimensional Ising spin-glass Hamiltonians.  One way to remedy this
situation is to increase the percolation threshold artificially, e.g.,
by diluting the lattice \cite{joerg:05}.  However, this is often not
desirable and is highly dependent on the problem to be studied.

Here, we show that Houdayer-like cluster moves can be applied to spin
systems on topologies where the percolation threshold is below 50\%,
provided that the interplay of temperature and frustration prevents
clusters from spanning the whole system.  We therefore introduce {\em
isoenergetic cluster moves} for spin-glass-like Hamiltonians in any
space dimension. These rejection-free cluster moves accelerate
thermalization by several orders of magnitude even for systems with
space dimensions larger than $2$. We show that the inherent frustration
present in spin-glass Hamiltonians prevents clusters from spanning the
whole system for temperatures below the characteristic energy scale of
the problem.  As such, spin-glass simulations can be sped up
considerably in the hard-to-reach low-temperature regime of interest in
many numerical studies.

The fact that the isoenergetic cluster moves are rejection free and
leave the energy of the system unchanged is also of great importance to
any heuristic based on Monte Carlo updates to compute ground-state
configurations of spin-glass-like Hamiltonians. For example, the
convergence of simulated annealing \cite{kirkpatrick:83} can be
considerably improved by adding isoenergetic cluster moves at each
temperature step. Because the moves change the spin configurations but
leave the energy of the system intact, the approach has the potential to
``tunnel'' through energy barriers, thus improving overall convergence.

We first introduce the benchmark model, followed by a short description
of the Houdayer cluster algorithm and an outline of our isoenergetic
cluster algorithm.  Results in two and three space dimensions, as well
as on the nonplanar chimera topology \cite{bunyk:14} are presented.

{\em Benchmark model and observables.---}The Hamiltonian of a generic
Ising spin glass is defined by $\mathcal{H} = \sum_{i \neq j}^N
J_{ij}s_{i}s_{j}$, where $s_i \in \lbrace \pm 1 \rbrace$ represent Ising
spins and $N$ is the total number of spins.  In this study the
interactions $J_{ij}$ are selected from a Gaussian distribution with
mean zero and variance $J^2 = 1$.  Because we are only interested in
highlighting the improved thermalization by adding isoenergetic cluster
moves, we measure the average energy per spin defined via $[\langle
{\mathcal H}\rangle]/N$, as well as the link overlap $q_{\ell} =
(1/N_b)\sum_{ij}^{N} s_i^{(1)} s_j^{(1)} s_i^{(2)} s_j^{(2)}$. Here,
$\langle \cdots \rangle$ represents a Monte Carlo average, the
superscripts represent two replicas of the system, $[\cdots]$ indicates
an average over the disorder, and $N_b$ is the number of bonds in the
system.  Using Gaussian disorder, one can equate the internal energy per
spin to the internal energy computed from the link overlap
\cite{katzgraber:01}, $E(q_\ell)$, i.e.,
\begin{equation}
E(q_{\ell}) = - \frac{J^2}{T} \frac{N_b}{N} (1 - q_{\ell}).
\end{equation}
To test that the system is thermalized, we thus study the time-dependent
behavior of 
\begin{equation}
\Delta = [ \langle E (q_{\ell}) \rangle - \langle {\mathcal H}/N \rangle ] .
\label{eq:delta}
\end{equation}
When $\Delta \to 0$, the bulk of the disorder instances are thermalized
\cite{yucesoy:13}.  Simulation parameters are listed in Table
\ref{tab:simparams}

{\em Reminder: Houdayer cluster algorithm.---}The Houdayer cluster
algorithm (HCA) \cite{houdayer:01} is an efficient algorithm to study
{\em two-dimensional} Ising spin glasses at low temperatures where
thermalization is slow. It is similar to replica Monte Carlo
\cite{swendsen:86}, but with the difference that both replicas are at
the {\em same} temperature.  By allowing large cluster rearrangements of
configurations, the HCA improves thermalization by efficiently tunneling
through configuration space.

The algorithm works as follows: In the HCA, two independent spin
configurations (replicas) are simulated at the same temperature.  The
site overlap between replicas $(1)$ and $(2)$, $q_i = s_i^{(1)}
s_i^{(2)}$, is calculated. This creates two domains in $q$ space: sites
with $q_i = 1$ and $q_i = -1$.  Clusters are defined as the connected
parts of these domains in $q$ space.  One then randomly chooses one site
with $q_i = -1$ and builds the cluster by adding all of the connected
spins in the domain with probability $1$. When no more spins can be
added to the cluster in $q$ space, the spins in {\em both} replicas that
correspond to cluster sites are flipped with probability $1$,
irrespective of their orientation. The method can be implemented in a
very efficient way because sites are added to the cluster with
probability $1$ and the cluster updates are rejection free.  To ensure
ergodicity, the cluster move is combined with standard single-spin Monte
Carlo updates. Summarizing, one simulation step using the HCA consists
of the following steps:

\begin{enumerate}

\item{Perform one Monte Carlo sweep ($N$ Metropolis updates) in each
replica.}

\item{Perform one Houdayer cluster move.}

\item{Perform one parallel tempering update for a pair of neighboring
temperatures}.

\end{enumerate}

Note that the last step is not necessary; however, the combination of
the HCA moves and parallel tempering (PT) updates improves
thermalization considerably and represents the standard {\em modus
operandi}.

\begin{figure}[]
\includegraphics[width=0.87\columnwidth]{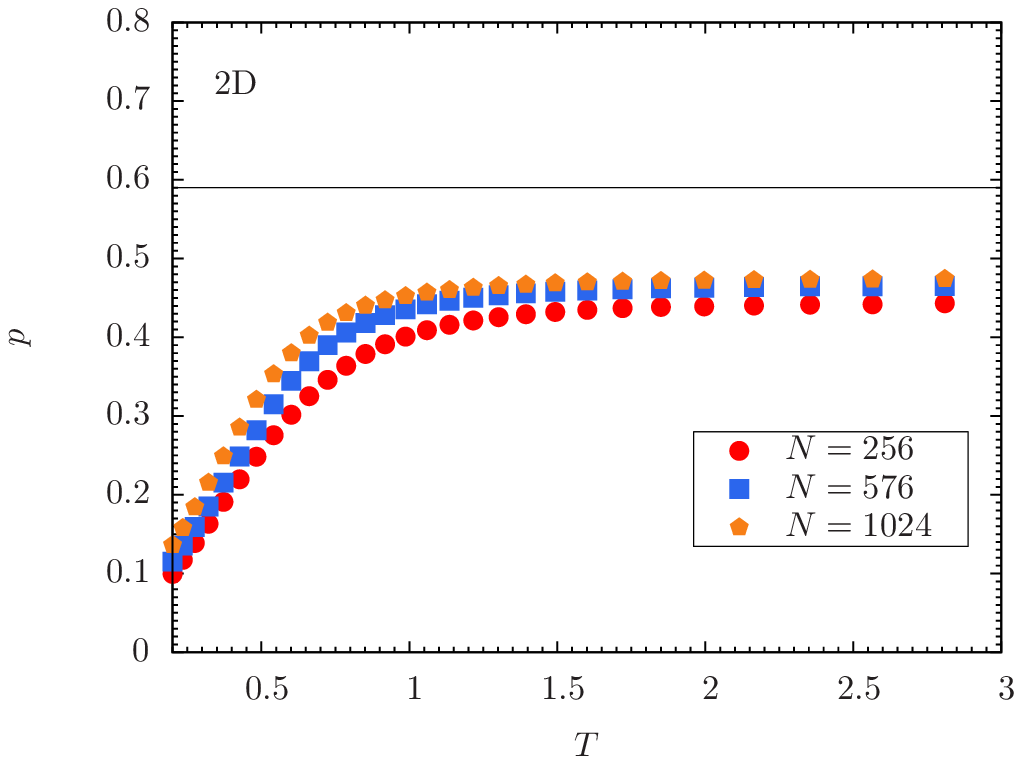}
\includegraphics[width=0.87\columnwidth]{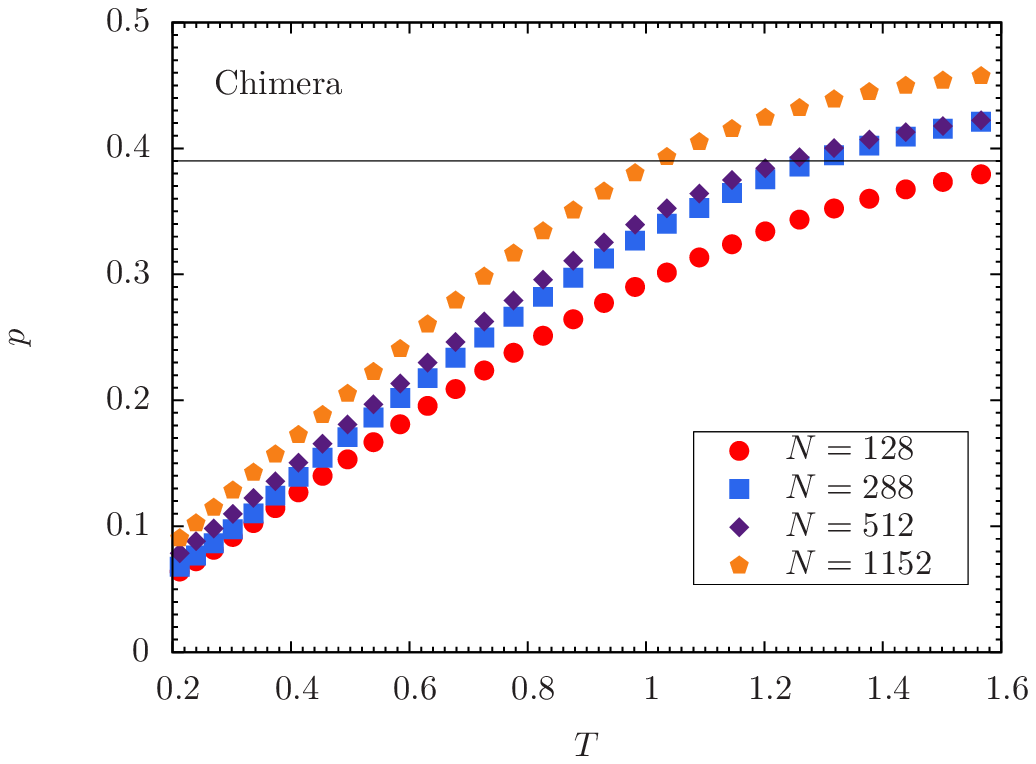}
\includegraphics[width=0.87\columnwidth]{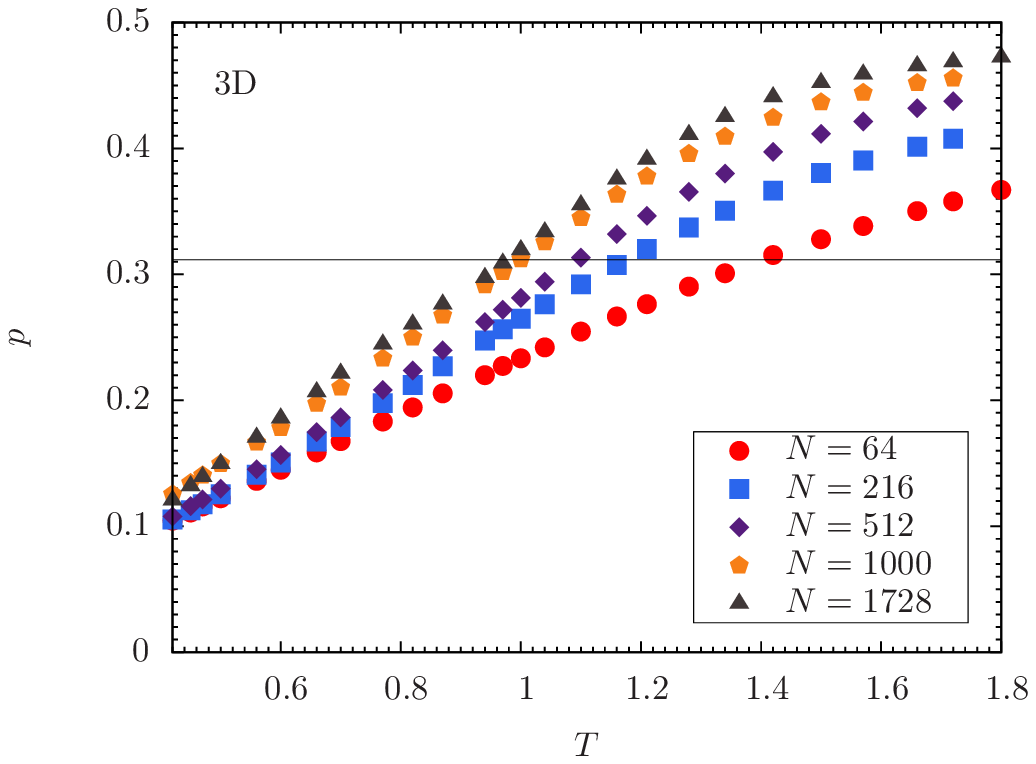}

\caption{(color online).
(Top panel) Fraction of spins $p$ of potential cluster sites as a
function of temperature $T$ for different system sizes $N$ in two space
dimensions (2D).  The horizontal line represents the percolation
threshold of a two-dimensional square lattice, i.e., $p_c \approx 0.592$
\cite{feng:08}.  Because $p \to 0.5$ for $T \to \infty$, for all $T$
clusters do not percolate, which is why the HCA is efficient in
two-dimensional planar geometries. (Center panel) $p$ as a function of
temperature $T$ for different system sizes $N$ on the chimera topology.
The horizontal line represents the percolation threshold of the
nonplanar chimera topology, namely $p_c \approx 0.387$ computed here
using the approach developed in Ref.~\onlinecite{melchert:13} (see the
Supplemental Material). For $T \gtrsim J = 1$ clusters percolate and
cluster updates provide no gain. (Bottom panel) $p$ as a function of
temperature $T$ for different system sizes $N$ in three space dimensions
(3D).  The horizontal line represents the percolation threshold of the
three-dimensional cubic lattice ($p_c \approx 0.311$ \cite{wang:13a}).
For $T \gtrsim J =1$ clusters percolate.  In all panels, error bars are
computed via a jackknife analysis over configurations and are smaller
than the symbols.
}
\label{fig:prob}
\end{figure}

In theory, the efficiency of the HCA depends strongly on the percolation
threshold of the desired topology to be simulated. Because spins are
added to the cluster with probability $1$, if the percolation threshold
of the studied lattice is below 50\%, then the cluster might span the
entire system and an update will not yield a new configuration. This is
the reason why the HCA is claimed to only work in two space dimensions
\cite{houdayer:01} where the percolation threshold is above 50\% (see
also Fig.~\ref{fig:prob}, top panel).

{\em Isoenergetic cluster algorithm.---}Our proposed isoenergetic
cluster moves are closely related to the HCA.  We begin by simulating
two replicas with the same disorder at multiple temperatures. The
cluster moves alone are not ergodic, so, again, these must be combined
with simple Monte Carlo updates.  One simulation step using isoenergetic
cluster moves consists of the following steps:

\begin{enumerate}

\item[1.]{Perform one Monte Carlo sweep ($N$ Metropolis updates) in each
replica.}

\item[2a.]{If the number of cluster sites with $q_i = -1$ is greater
than $N/2$, then all the spins in one of the configurations can be
flipped (because of spin-reversal symmetry), thus reducing the cluster
size while leaving the energy unchanged.}

\item[2b.]{Perform one Houdayer cluster move for all temperatures $T
\lesssim J$.}

\item[3.]{Perform one parallel tempering update for a pair of neighboring
temperatures}.

\end{enumerate}

The main difference thus lies in applying cluster moves to a carefully
selected set of temperatures where the isoenergetic cluster moves (ICMs)
are efficient (steps 2a and 2b) because clusters do not percolate, as
well as reducing cluster sizes and thus the numerical overhead by
exploiting spin-reversal symmetry (step 2a)
\cite{comment:FM,comment:implementation}. For example, in the case of
the chimera lattice the overhead of the ICM over PT is approximately
25\% and is roughly independent of the system size for the studied $N$.
However, the overhead for the HCA over PT is at least 50\% and grows
with increasing system size.

Figure \ref{fig:prob} shows the fraction of spins with negative overlap
(i.e., the fraction of potential cluster sites) as a function of
temperature $T$ for different system sizes $N$ and on three different
topologies. The top panel of Fig.~\ref{fig:prob} shows data in two space
dimensions where the percolation threshold is $p_c \approx 0.592$
\cite{feng:08} (the solid horizontal line). As such, for all
temperatures simulated, the fraction of cluster sites is below the
percolation threshold and saturates at 50\% for $T \to \infty$. This
means that isoenergetic cluster updates are efficient for all
temperatures studied because the clusters never percolate. Naively, one
would expect that in higher space dimensions clusters percolate for all
$T$'s. This is, however, not the case due to the frustration present in
spin glasses, as can be seen for the chimera topology (the center panel
of Fig.~\ref{fig:prob}) or in three space dimensions (the bottom panel
of Fig.~\ref{fig:prob}). For increasing system size the fraction of
cluster sites converges to a limiting curve that crosses the percolation
threshold (the horizontal solid lines) at approximately $T \approx J  =
1$.  This means that, for all $T \gtrsim J$, clusters percolate and the
cluster updates are just numerical overhead without any advantage to the
simulation. However, for $T \lesssim J$ the fraction of cluster sites
lies below the percolation threshold.  This means that performing
cluster moves in this temperature regime should improve thermalization.
Note that it is a coincidental property that for three-dimensional Ising
spin glasses $T_c \sim 1$ \cite{katzgraber:06}, i.e., that cluster moves
can be applied to any $T \lesssim T_c$ \cite{comment:4D}.

When the interactions $J_{ij}$ are drawn from a Gaussian distribution,
the ground state is unique. As can be seen in Fig.~\ref{fig:prob}, the
fraction $p$ of spins potentially in a cluster also approaches zero for
$T \to 0$; i.e., both replicas are in the ground state for low enough
$T$.  Therefore, the cluster is composed of no sites or the entire
lattice.  In the case of disorder distributions that yield a highly
degenerate ground state, such as is the case for bimodal disorder, it is
possible to continue to have clusters at zero temperature. It is thus
possible to efficiently hop around the ground-state manifold by applying
cluster moves to low-lying or even zero-temperature states, although
this might not be ergodic.  We do emphasize, however, that if clusters
are too small, then the isoenergetic cluster moves also become
ineffective.  Therefore, plotting $p$ as was done in Fig.~\ref{fig:prob}
is essential in determining the efficiency and applicability of the
method.

\begin{table}[h]
\caption{
Parameters of the simulation in two space dimensions (2D), three space
dimensions (3D), and on the chimera (Ch) topology.  For each topology
simulated and system sizes $N$, we compute $N_{\rm sa}$ disorder
instances and measure over $2^b$ Monte Carlo sweeps (and isoenergetic
cluster moves) for each of the $2 N_T$ replicas. $T_{\rm min}$ [$T_{\rm
max}$] is the lowest [highest] temperature simulated, and $N_T$ is the
total number of temperatures used in the parallel tempering Monte Carlo
method. Isoenergetic cluster moves only occur for the lowest $N_{\rm c}$
temperatures simulated (determined from Fig.~\ref{fig:prob}).
\label{tab:simparams}
}
\begin{tabular*}{\columnwidth}{@{\extracolsep{\fill}} l l l l l l l r }
\hline
\hline
& $N$ & $N_{\rm sa}$ & $b$ & $T_{\rm min}$ & $T_{\rm max}$ & $N_{T}$ &$N_{\rm c}$  \\
\hline
2D & $256$, $576$, $1024$ & $10^4$ 
	& $22$ & $0.2120$ & $1.6325$ & $30$ &$30$ \\
Ch & $128$, $288$, $512$, $800$, $1152$  
	& $10^4$ & $22$ & $0.2120$ & $1.6325$ & $30$ &$19$ \\
3D & $64$, $216$, $512$, $1000$, $1728$   
	& $1.5 \, 10^4$ & $23$ & $0.4200$ & $1.8000$ & $26$ &$13$\\
\hline
\hline
\end{tabular*}
\end{table}

\begin{figure}[]
\includegraphics[width=0.87\columnwidth]{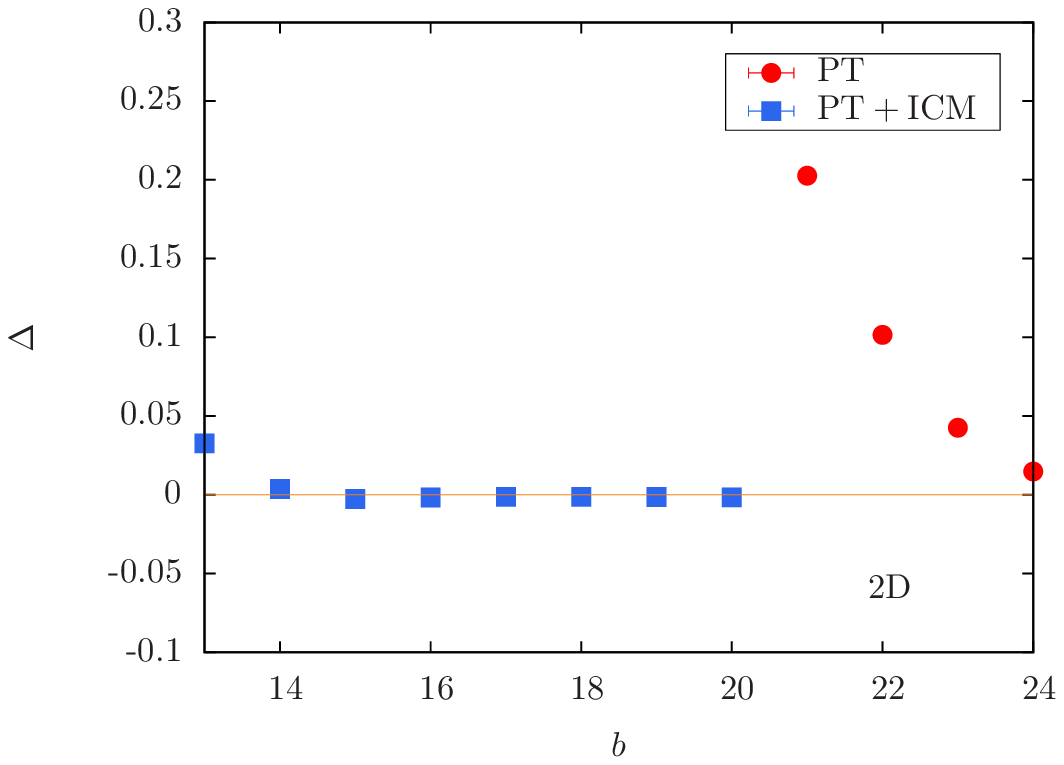}
\includegraphics[width=0.87\columnwidth]{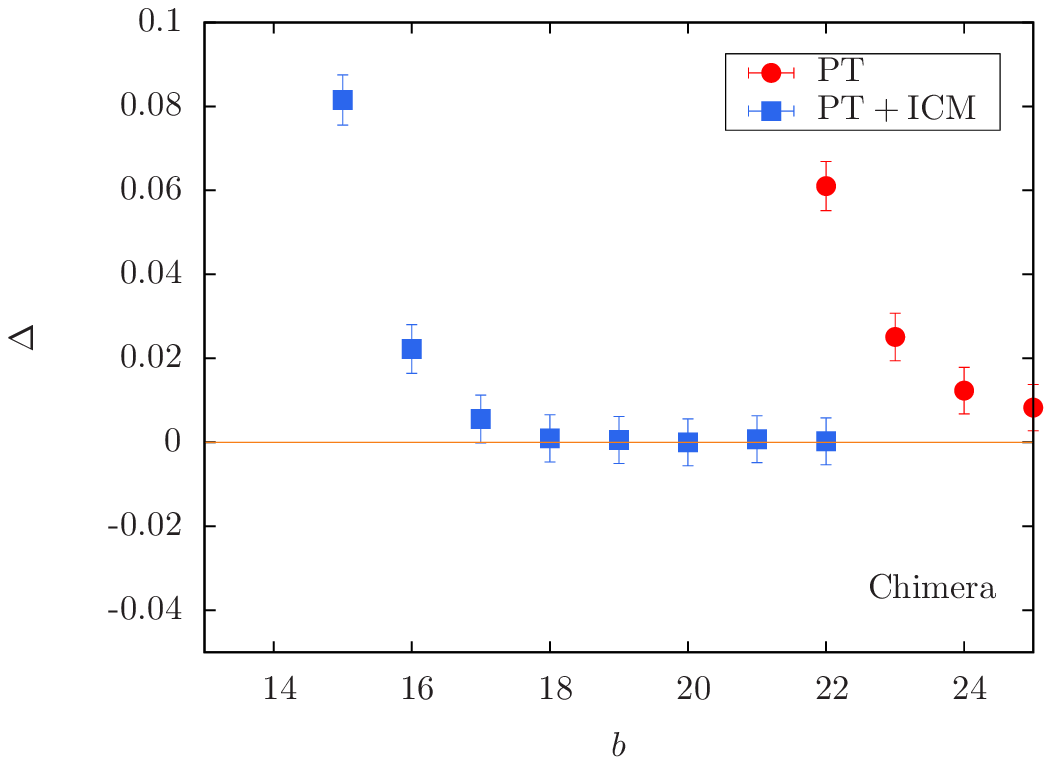}
\includegraphics[width=0.87\columnwidth]{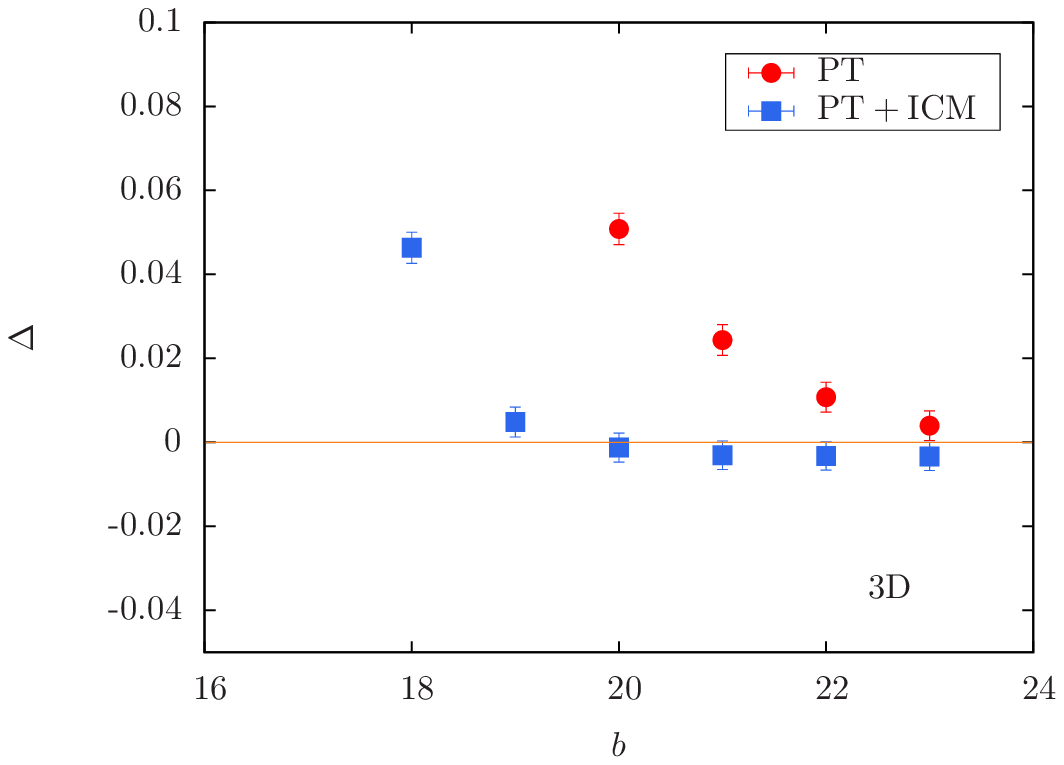}

\caption{(color online).
(Top panel) $\Delta$ [Eq.~\eqref{eq:delta}] as a function of simulation
time $t = 2^b$ measured in Monte Carlo sweeps in two space dimensions
(2D) for $N = 1024$ and $T = 0.212$.  Simulations using vanilla PT
thermalize at at least $2^{25}$ Monte Carlo sweeps, whereas with
the addition of ICMs thermalization is reduced to approximately $2^{16}$
Monte Carlo sweeps. This means approximately 2 orders of magnitude
improvement. (Center panel) $\Delta$ as a function of simulation time $t
= 2^b$ measured in Monte Carlo sweeps for an Ising spin glass on chimera
with $N = 1152$ spins at $T = 0.212$.  Simulations using PT thermalize
at approximately $2^{25}$ Monte Carlo sweeps, whereas the addition of
ICMs reduces thermalization to $2^{18}$ Monte Carlo sweeps.  Again,
approximately 2 orders of magnitude speedup. (Bottom panel) $\Delta$
as a function of simulation time $t = 2^b$ measured in Monte Carlo
sweeps in three space dimensions (3D) for $N = 1728$ and $T = 0.42 \sim
0.43T_c$. Using standard PT, the system thermalizes approximately after
$2^{23}$ Monte Carlo sweeps.  This time is reduced to $\sim 2^{20}$
Monte Carlo sweeps when ICMs are added.  In all panels, error bars are
computed via a jackknife analysis over configurations.
}
\label{fig:therm}
\end{figure}

{\em Benchmarking results.---}Figure \ref{fig:therm} shows $\Delta$
[Eq.~\eqref{eq:delta}] as a function of Monte Carlo time (measured in
lattice sweeps) $t = 2^b$. The top panel of Fig.~\ref{fig:therm} shows
data in two space dimensions for simulations using isoenergetic cluster
moves (PT+ICM) and vanilla PT Monte Carlo for $N = 1024$ spins at $T =
0.212$. Once $\Delta \sim 0$, we deem the system thermalized. Clearly,
the inclusion of cluster moves---as can also be expected from the
results of Houdayer---show an improved thermalization.  The center panel
of Fig.~\ref{fig:therm} shows data on the chimera topology with $N =
1152$ spins and $T = 0.212$, where the HCA is not expected to show any
improvement over PT due to $p_c < 0.5$. As can be seen, our ICM clearly
improve thermalization in comparison to PT by at least 2 orders of
magnitude, an amount that grows with increasing system size.  Finally,
the bottom panel of Fig.~\ref{fig:therm} shows $\Delta$ as a function of
simulation time in three space dimensions with $N = 1728$ spins and $T =
0.42 \ll T_c$.  Although not as impressive as with the chimera topology,
we see a speedup of approximately one order of magnitude---an amount
that again grows with increasing system size.

\begin{figure}[h]
\includegraphics[width=0.90\columnwidth]{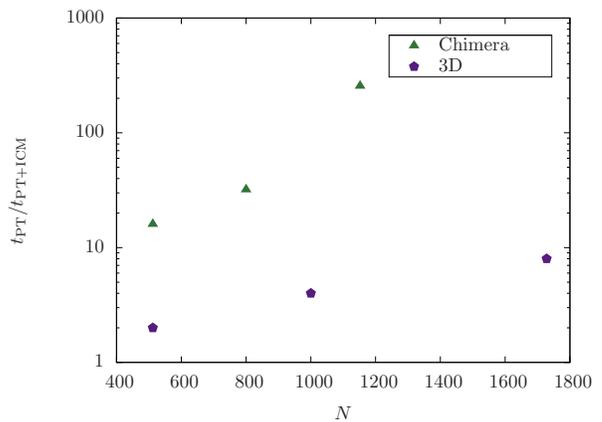}
\caption{(color online).
Ratio between the {\em approximate} average thermalization time of PT
and PT+ICM for different topologies at the lowest simulation temperature
(see Table \ref{tab:simparams}) as a function of system size $N$.  In all
cases the speedup increases with increasing system size. Note that
thermalization times have been determined by eye.
}
\label{fig:ratio}
\end{figure}

Finally, Fig.~\ref{fig:ratio} shows the ratio of the thermalization time
using PT and using PT+ICM for different topologies at the lowest
simulation temperature (see Table \ref{tab:simparams}) as a function of
the system size $N$. In all cases, the speedup increases with increasing
system size, therefore illustrating that the addition of isoenergetic
cluster moves greatly improves thermalization.

{\em Summary.---}We have presented a rejection-free cluster algorithm
for spin glasses in any space dimension that greatly improves
thermalization.  By restricting Houdayer cluster moves to temperatures
where cluster percolation is hampered by the interplay of frustration
and temperature, we are able to extend the Houdayer cluster algorithm
for two-dimensional spin glasses to any topology or space dimension. Our
standard implementation of the cluster updates represents only a minor
overhead \cite{comment:implementation} compared to the thermalization
time speedup obtained from the isoenergetic cluster moves---a speedup
that increases with the system size \cite{comment:tauto}.

We would like to thank F.~Hamze, J.~Machta and M.~Weigel for the
fruitful discussions.  H.~G.~K.~acknowledges support from the NSF (Grant
No.~DMR-1151387) and would like to thank Suntory's Hibiki and Yamazaki
for inspiration.  We thank Texas A\&M University for the extensive CPU
time on the Ada cluster.  This research is based upon work supported in
part by the Office of the Director of National Intelligence (ODNI),
Intelligence Advanced Research Projects Activity (IARPA), via MIT
Lincoln Laboratory Air Force Contract No.~FA8721-05-C-0002.  The views
and conclusions contained herein are those of the authors and should not
be interpreted as necessarily representing the official policies or
endorsements, either expressed or implied, of ODNI, IARPA, or the
U.S.~Government.  The U.S.~Government is authorized to reproduce and
distribute reprints for Governmental purpose.

\vspace*{1.0em}
\hrule
\vspace*{1.0em}

\begin{figure}[h]
\includegraphics[width=0.90\columnwidth]{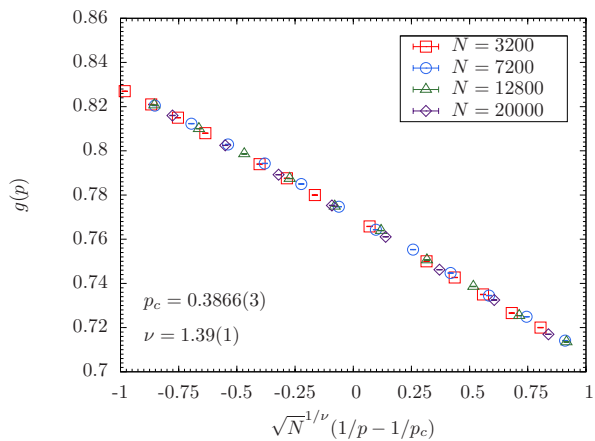}
\caption{(Color online)
Finite-size scaling of the percolation probability Binder
ratio $g$ for different lattice sizes $N$ on the chimera topology.
Best scaling is obtained for $p_c = 0.3866(3)$ and $\nu = 1.39(1).$
}
\label{fig:perc}
\end{figure}

{\em Supplemental Material: Percolation threshold of the chimera
lattice.---}In percolation theory, the relative size of the largest
cluster of a lattice $\langle s_{\rm max}\rangle$ plays the role of an
order parameter. Above the percolation threshold $p_c$, $\langle s_{\rm
max}\rangle \to 1$, whereas below $p_c$ it approaches $0$. One can
define a dimensionless Binder ratio \cite{melchert:13} via $g =
(1/2)[3-(\langle s_{\rm max}^4\rangle/\langle s_{\rm max}^2\rangle^2)]$
to determine the percolation threshold $p_c$ to high precision. Here,
$\langle \cdots \rangle$ represents an average over randomly-generated
configurations with a site probability $p$ on a chimera lattice with $N$
sites. Close to criticality, $g \sim G[(\sqrt{N})^{1/\nu}(1/p -
1/p_c)]$, i.e., when $p = p_c$ data for different system sizes cross.
Figure \ref{fig:perc} shows a finite-size scaling plot of the data for
$g$. We estimate for the percolation threshold of the chimera lattice
$p_c = 0.3866(3)$.

\bibliography{refs,comments}

\begin{thebibliography}{25}
\bibitem[{com({\natexlab{a}})}]{comment:FM}
\bibinfo{note}{Jon Machta tells us that the method might also work for
  ferromagnetic systems at low temperatures. Here, the step of flipping the
  whole spin configuration of one replica if the two replicas are in opposite
  pure states is crucial. In this case both replicas will be in the same pure
  state (all spins up or all spins down), and the clusters will be small in a
  ferromagnet as well.}
\bibitem[{com({\natexlab{b}})}]{comment:implementation}
\bibinfo{note}{Note that in step (2a) of the ICM algorithm we flip the signs of
  the spins to reduce the overhead whenever the number of cluster sites exceeds
  $N/2$. To ensure that the measurement of observables is not affected by this
  operation, the signs of the flipped spins need to be kept track of.}
\bibitem[{com({\natexlab{c}})}]{comment:4D}
\bibinfo{note}{Note that for the three-dimensional case, one may be tempted to
  think that clusters do not percolate because the spins are ``frozen'' below
  the critical temperature $T_c \sim 1$. However, this is not the case. In four
  space dimensions, $T_c \sim 1.8$ \cite{joerg:08d} and $p_c = 0.197$
  \cite{paul:01}. However, clusters still only stop percolating for $T \lesssim
  1$.}
\bibitem[{com({\natexlab{d}})}]{comment:tauto}
\bibinfo{note}{We note that our ICMs also slightly reduce autocorrelation times
  in addition to greatly improving thermalization times. However, note that the
  error due to configurational averages dominates in spin-glass simulations.}
\end{thebibliography}

\end{document}